
\input harvmac
%
%
%
%
\ifx\answ\bigans
\else
\output={
  \almostshipout{\leftline{\vbox{\pagebody\makefootline}}}\advancepageno
}
\fi
%
%
%

%
%

%
%
\def\UCSD#1#2{\noindent#1\hfill #2%
\bigskip\supereject\global\hsize=\hsbody%
\footline={\hss\tenrm\folio\hss}}
%
%
\def\abstract#1{\centerline{\bf Abstract}\nobreak\medskip\nobreak\par #1}
%
%
%
%
\edef\tfontsize{ scaled\magstep3}
 \tfontsize  \tfontsize
 \tfontsize \font\titlei=cmmi10 \tfontsize
\font\titleis=cmmi7 \tfontsize \font\titleiss=cmmi5 \tfontsize
\font\titlesy=cmsy10 \tfontsize \font\titlesys=cmsy7 \tfontsize
\font\titlesyss=cmsy5 \tfontsize  \tfontsize
\skewchar\titlei='177 \skewchar\titleis='177 \skewchar\titleiss='177
\skewchar\titlesy='60 \skewchar\titlesys='60 \skewchar\titlesyss='60
%
%
%
%
%
\def\inv{^{\raise.15ex\hbox{${\scriptscriptstyle -}$}\kern-.05em 1}}
\def\lbar{{\lower.35ex\hbox{$\mathchar'26$}\mkern-10mu\lambda}}
\def\e#1{{\rm e}^{^{\textstyle#1}}}

%
%
%
%
\def\slash#1{\rlap{$#1$}/} 
\def\dsl{\,\raise.15ex\hbox{/}\mkern-13.5mu D}
\def\delsl{\raise.15ex\hbox{/}\kern-.57em\partial}
\def\Ksl{\hbox{/\kern-.6000em\rm K}}
\def\Asl{\hbox{/\kern-.6500em \rm A}}
\def\Dsl{\hbox{/\kern-.6000em\rm D}} 
\def\Qsl{\hbox{/\kern-.6000em\rm Q}}
\def\gradsl{\hbox{/\kern-.6500em$\nabla$}}
%
%
\def\lspace{\ifx\answ\bigans{}\else\qquad\fi}
\def\lbspace{\ifx\answ\bigans{}\else\hskip-.2in\fi} 
%
%
\def\boxeqn#1{\vcenter{\vbox{\hrule\hbox{\vrule\kern3pt\vbox{\kern3pt
        \hbox{${\displaystyle #1}$}\kern3pt}\kern3pt\vrule}\hrule}}}
%
%
\def\mbox#1#2{\vcenter{\hrule \hbox{\vrule height#2in
\kern#1in \vrule} \hrule}}
%
%
%
%
   
   \def\CH{{\cal H}}
   
\def\CM{{\cal M}}  \def\CO{{\cal O}}

%
%
%
%
%

%

\def\bar#1{\overline{#1}}

\def\darr#1{\raise1.5ex\hbox{$\leftrightarrow$}\mkern-16.5mu #1}

%
%
\def\frac#1#2{{\textstyle{#1\over #2}}} 
%
%
%
%

%
%
%
%

%
%
\def\ltap{\ \raise.3ex\hbox{$<$\kern-.75em\lower1ex\hbox{$\sim$}}\ }
\def\gtap{\ \raise.3ex\hbox{$>$\kern-.75em\lower1ex\hbox{$\sim$}}\ }
\def\gl{\ \raise.5ex\hbox{$>$}\kern-.8em\lower.5ex\hbox{$<$}\ }
\def\roughly#1{\raise.3ex\hbox{$#1$\kern-.75em\lower1ex\hbox{$\sim$}}}
%
%

\def\etal{\hbox{\it et al.}}

\relax

\def\bbar{\overline {\rm b}}
\def\sbar{\overline {\rm s}}
\def\gone{\Gamma_1}
\def\gtwo{\Gamma_2}
\def\hbar{\bar h_Q}

\def\qhat{\hat q}
\def\ms{m_s}
\def\mshat{\hat\ms}
\def\qhsl{\slash{\qhat}}
\def\vsl{\slash{v}}
\def\qsl{\hbox{/\kern-.5600em {$q$}}}
\def\ksl{\hbox{/\kern-.5600em {$k$}}}
\def\id{{\rm iD}}
\def\({\left(}
\def\){\right)}
\def\dtk{ {{\rm d}^3 k\over (2\pi)^3 2 E_\gamma} }

\def\vdotq{v\cdot\qhat}
\def\lamone{\lambda_1}
\def\lamtwo{\lambda_2}
\def\bsg{{B\rightarrow X_s\gamma}}
\def\bsee{{B\rightarrow X_s\ell^+\ell^-}}
\def\gamleft{\frac12(1-\gamma_5)}
\def\gamright{\frac12(1+\gamma_5)}

\def\km{V_{tb} V_{ts}^*}

\def\lqcd{\Lambda_{\rm QCD}}
\def\shat{\hat s}
\def\shatsq{\hat s^2}
\def\b{{\rm b}}
\def\s{{\rm s}}
\def\c{{\rm c}}
\def\u{{\rm u}}
\def\e{{\rm e}}
\def\O{{\cal O}}
\def\L{{\cal L}}
\def\d{{\rm d}}
\def\D{{\rm D}}
\def\im{{\rm i}}
\def\q{{\rm q}}
\def\vslash{v\hskip-0.5em /}
\def\Dslash{{\rm D}\hskip-0.7em /\hskip0.2em}

\def\gev{\,{\rm GeV}}
\def\ol{\overline}
\def\OMIT#1{}
\def\frac#1#2{{#1\over#2}}
\def\lamqcd{\Lambda_{\rm QCD}}
\def\etal{{\it et al.}}
\def\coeff#1#2{{\textstyle {#1\over#2}}}

\hbadness=10000

\noblackbox
\vskip 1.in
\centerline{{\titlefont{Nonperturbative Contributions to the
Inclusive}}}
\medskip
\centerline{{\titlefont{Rare Decays $B\to X_s\gamma$ and
$B\to X_s\ell^+\ell^-$}
\footnote{*}{{\tenrm Work
supported in part by the Department of Energy under contracts
DE--AC03--76SF00515 (SLAC), DE--FG03--90ER40546 (UC San Diego),
DE--FG02--91ER40682 (CMU).}}}}
\vskip .5in
\centerline{Adam F.~Falk${}^{abc}$, Michael Luke${}^{ad}$ and
Martin J.~Savage${}^{ae}$}
\medskip
{\it{
\centerline{a) Department of Physics, University of California, San
Diego, La Jolla CA 92037}
\centerline{b) Stanford Linear Accelerator Center, Stanford CA 94309}
\centerline{c) Department of Physics and Astronomy, The Johns Hopkins
University,}
\centerline{Baltimore MD 21218}
\centerline{d) Department of Physics, University of Toronto, Toronto,
Canada M5S 1A7}
\centerline{e) Department of Physics, Carnegie Mellon University,
Pittsburgh PA 15213}}}

\vskip .2in

\abstract{
We discuss nonperturbative contributions to the inclusive rare $B$
decays $B\to X_s\gamma$ and $B\to X_s\ell^+\ell^-$.  We employ an
operator product expansion and the heavy quark effective theory to
compute the leading corrections to the decay rate found in the free
quark decay model, which is exact in the limit $m_b\to\infty$.  These
corrections are of relative order $1/m_b^2$, and may be parameterised
in terms of two low-energy parameters.  We also discuss the corrections
to other observables, such as the average photon energy
in $B\to X_s\gamma$ and the lepton invariant mass spectrum
in $B\to X_s\ell^+\ell^-$.}

\vfill
\UCSD{\vbox{
\hbox{UCSD/PTH 93-23}
\hbox{SLAC-PUB-6317}
\hbox{JHU-TIPAC-930020}}\hskip .2in
\vbox{\hbox{UTPT 93-19}
\hbox{CMU-HEP 93-12}
\hbox{DOE-ER/40682-37}}
}{August 1993}
\eject

\newsec{Introduction}

The rare decays of $B$ mesons have never been of greater interest, both
experimentally and theoretically.  The first observation of a decay
mediated by the quark transition $\b\to\s$ recently has been reported
by the CLEO Collaboration \ref\cleo{R. Ammar \etal, {\sl Phys. Rev.
Lett.} 71 (1993) 674}, who found a branching fraction for the process
$B\to K^*\gamma$ of $(4.5\pm1.9\pm0.9)\times 10^{-5}$.  Such
transitions are typically induced by the exchange of virtual heavy
quanta, the effects of which appear at low energies as local operators
multiplied by small coefficients.  It is hoped that the detection of
these suppessed interactions in the guise of rare $B$ decays may
provide a direct window to physics at much higher scales.

In order for such a hope to be realised, however, it is necessary to
connect the quark-level operators which are generated perturbatively to
the hadronic transitions which are actually observed.  This involves
the consideration of nonperturbative hadronic matrix elements, which
typically are incalculable.  One common approach to this problem is to
consider inclusive rates such as $B\to X_s$ rather than individual
exclusive channels, and to model the inclusive transition by the decay
of a free bottom quark to a free strange quark.  It is hoped that for
the b quark mass $m_b$ sufficiently large, the operator mediating
$\b\to\s$ acts over distances short compared to the scales of
confinement and strong QCD interactions, and the approximation is a
good one.

The issue of how good this approximation really is originally was
addressed by Chay, Georgi and Grinstein \ref\chayetal{J. Chay, H.
Georgi and B. Grinstein, {\sl Phys. Lett.} B247 (990) 399}.  Using the
tools of the heavy quark effective theory (HQET), they showed that the
free quark model is in fact the first term in a controlled expansion in
$1/m_b$, and hence is arbitrarily accurate as $m_b\to\infty$.  In
addition, they demonstrated that there are no contributions to the
rate at subleading ($1/m_b$) order, and that any corrections could only
come in at order $1/m_b^2$ or higher.

In this paper, we extend the work of Chay {\it et al.}~to compute the
leading corrections to free quark decay, for the inclusive processes
$B\to X_s\gamma$ and $B\to X_s\ell^+\ell^-$.  While the $1/m_b^2$
corrections here are not particularly large, it is important to know
their size if the free quark decay model is to be trusted.  We also
believe that our computation is a very nontrivial application of HQET
in a somewhat unfamiliar regime, and is hence quite interesting in its
own right.

Finally, we note that work which overlaps with ours has been performed
recently, in a somewhat different formalism, by Bigi {\it et
al.}~\ref\russians{I.I. Bigi, N. G. Uraltsev and A. I. Vainshtein, {\sl
Phys. Lett.} B293 (1992) 430; I.I. Bigi, B. Blok, M. Shifman, N.G.
Uraltsev and A. Vainshtein, TPI-MINN-92/67-T (1992)}.

\newsec{The Operator Product Expansion and Matrix Elements in HQET}

In this section we will discuss our procedure in general terms, to
elucidate the structure of the expansion before beseiging the reader
with particular details.  We are interested in the rare decays of b
quarks, such as $\b\to\s\gamma$ or $\b\to\s\e^+\e^-$, which are
mediated at low energies by local operators of the form
\eqn\typo{
    \O P(\phi)=\ol\s\Gamma\b\,P(\phi)\,.}
Here $P(\phi)$ is meant to stand for some function of perturbatively
interacting fields such as leptons or a photon, and $\Gamma$ is a
general Dirac structure.  Interactions such as \typo\ are typically
induced at high energies by the exchange of virtual $W$ bosons, top
quarks, or new exotic quanta.  At low energies they appear in the
effective Hamiltonian as local operators, with coefficients which may
be computed using renormalisation group techniques.  We will take the
presence of such operators simply as given; our interest will be in the
evaluation of their hadronic matrix elements.  We note that operators
of the form \typo\ are not the only relevant ones which will appear at
low energies; for example, we will typically find four-quark operators
as well.  For these, the techniques which we will present below will
only be appropriate when the invariant mass of the intermediate $q\bar
q$ pair is far from any quarkonium resonances. We will return to this
issue in our discussion of the decay $\b\to\s\e^+\e^-$.

For now, however, we restrict ouselves to operators with the structure
\typo. They induce quark level transitions of the form $\b\to\s$.
However, since the quarks are confined, what is observed is the decay
$B\to X_s$, in which a $B$ meson decays to an arbitrary hadronic state
$X_s$ with strangeness $S=-1$.  (Decays from the lowest lying bottom
baryon, $\Lambda_b$, are also possible.) Hence  we need to compute
matrix elements of the form
\eqn\matel{
    \langle X_s\cdots|\,\O P(\phi)\,|B\rangle\,,}
where the ellipses denote denote the additional perturbatively
interacting fields which couple to $P(\phi)$.  Unfortunately, exclusive
hadronic matrix elements such as \matel\ are governed by
nonperturbative strong interactions and are typically incalculable. At
best, $SU(3)$ and heavy quark symmetries may be used to relate the form
factors which appear in one such matrix element to those which appear
in another \ref\suthree{G. Burdman and J.F. Donoghue, {\sl Phys. Lett.}
B280 (1992) 287; A.F. Falk and B. Grinstein, SLAC--PUB--6237 (1993); R.
Casalbuoni \etal, UGVA-DPT 1993/04-816 (1993)}.  But computations from
first principles are not at this point possible.

Instead of considering the exclusive modes individually, then, we will
we will sum over all possible strange final states $X_s$. As has been
shown by Chay {\it et al.}~\chayetal, the inclusive decay rate may in
fact be calculated reliably. Previous computations of the inclusive
rate have relied on the free quark decay model, in which the sum over
exclusive decays is modeled by the decay of an on-shell bottom quark to
an on-shell strange quark. For $m_b\to\infty$ this is justified by
arguing that the decay is essentially a short distance process, which
occurs on time scales much shorter than those which govern the eventual
hadronisation of the final state.  This argument can be made precise
within a controlled expansion in inverse powers of the bottom quark
mass $m_b$ \chayetal, and we will be able to compute the leading
corrections to this limit.

Squaring the matrix element \matel\ and summing over $X_s$, we find a
differential decay rate of the form
\eqn\wdef{
    \d\Gamma={1\over2M_B}\sum_{X_s}\d[{\rm P.S.}]
    (2\pi)^4\delta^{(4)}(P_B-P_{X}-q)
    \langle B|\,\im\O^\dagger P(\phi)^\dagger\,|X_s\cdots\rangle\,
    \langle X_s\cdots|\,\im\O P(\phi)\,|B\rangle\,.}
Here $P_B$ and $P_X$ are the momenta of the intial $B$ and final $X_s$
systems, and $q=P_B-P_X$ is the momentum transfered to the other decay
products.  The symbol $\d[{\rm P.S.}]$ denotes an appropriate phase
space differential.  The part of $\d\Gamma$ which involves the fields
$P(\phi)$ may be calculated perturbatively.  We then find that
$\d\Gamma$ is equal to the product of known factors times an expression
$W(q)$ which involves only the quark and gluon fields:
\eqn\wdefii{
    W(q)=\sum_{X_s}(2\pi)^4\delta^{(4)}(P_B-P_{X}-q)\langle
    B|\,\O^\dagger\,|X_s\rangle\,\langle X_s|\,\O\,|B\rangle\,.}
Here the sum over $X_s$ includes the hadronic phase space integral. The
treatment of this nonperturbative expression is the subject of the rest
of this section.

We begin by noting that $W(q)$, being essentially a total decay rate,
is related by the optical theorem to the discontinuity in a forward
scattering amplitude.  That is, we may write
\eqn\wfort{
    W(q)=2{\rm Im}\,T(q)\,,}
where an example of the time-ordered product
\eqn\tdef{
    T(q)=\langle B|\,T\{\O^{\dagger},\O\}|B\rangle}
is shown in \fig\ope{Feynman diagrams contributing to the
time-ordered product $T\big\{\bbar\gone\s,\,\sbar\gtwo\b\big\}$.}.

Now we come to a crucial observation \chayetal. The sum over $X_s$ in
eq.~\wdefii\ includes hadronic states with a large range of invariant
masses, $M_K^2\le P_X^2\le M_B^2$.  The energy which flows into the
hadronic system $X_s$ scales with $m_b$ as the bottom mass increases,
and in the limit $m_b\to\infty$ is typically much larger than the
energy scale $\lamqcd$ which characterizes the strong interactions.
Hence, in all but a corner of the Dalitz plot, in which $P_X^2\approx
m_s^2$, the strange quark in \ope\ is far from its mass shell.  In
position space, this means that the points at which $\O$ and
$\O^\dagger$ act must be very near each other on the scale of
nonperturbative QCD, and it is appropriate to perform an operator
product expansion of the time-ordered product in eq.~\tdef. This
operator product expansion may be computed perturbatively in
$\alpha_s(m_b)$.  It will be valid over almost all of the Dalitz plot,
failing only in the region where $P_X^2$ is small.  In the large $m_b$
limit, the fractional contribution of this bad region to the total
phase space integral is negligible, and our calculation of the
inclusive decay rate based on this expansion will be reliable.  Our
approach, then, will be to perform a systematic expansion in inverse
powers of $m_b$, of which the leading term will be the result in the
$m_b\to\infty$ limit of the theory \chayetal. However, we will also be
able to compute the leading corrections to this limit, using the tools
of the heavy quark effective theory.

In this section we will discuss the form of the operator product
expansion, and how to take the hadronic matrix elements of the
operators which come out of it.  When we apply this formalism in the
following sections, the expressions which we derive sometimes will be
quite lengthy.  Here we will concentrate only on the structure of the
procedure. In general, then, the time-ordered product \tdef\ may be
expanded in a series of local operators suppressed by powers of the
mass of the bottom quark,
\eqn\opprod{
    T\{\O^{\dagger},\O\}\mathrel{\mathop=^{\rm OPE}}
    {1\over m_b}\left[ \O_0+{1\over2m_b}\O_1
    +{1\over4m_b^2}\O_2+\ldots\right]\,.}
The operator $\O_n$ is an operator of dimension $3+n$, with $n$
derivatives.

At this point, it is useful to introduce the heavy quark effective
theory (HQET) \ref\hqet{H. Georgi, {\sl Phys. Lett.} B240 (1990) 247;
E. Eichten and B. Hill, {\sl Phys. Lett.} B234 (1990) 511}, an
effective theory of QCD in which the mass of the b quark is taken to
infinity.  This effective theory implements on the lagrangian level the
new ``spin-flavor'' symmetry of QCD which arises in this limit
\ref\isgurwise{N. Isgur and M.B. Wise, {\sl Phys. Lett.} B232 (1989)
113; {\sl Phys. Lett.} B237 (1990) 527}.  Both
the mass and the spin of the b quark decouple from the soft bound state
dynamics of the hadron of which it is a part; so far as the light
degrees of freedom are concerned, the heavy quark is nothing but a
static, point-like source of color.  The exchange of soft gluons with
the light degrees of freedom leave the b quark always almost on shell.
Thus we can write its four-momentum $p_b^\mu$ as the sum of its
``on-shell'' momentum $m_bv^\mu$ and a ``residual momentum'' $k^\mu$,
such that the components of $k^\mu$ are always small compared to $m_b$.
It is then convenient to replace the usual quark field $\b(x)$ by a new
two-component field $h(x)$ with fixed four-velocity $v^\mu$,
\eqn\hintro{
    h(x)=\e^{\im m_bv\cdot x}P_+\b(x)\,,}
where $P_+=\frac12(1+\vslash)$ projects onto the quark, rather than
antiquark, degrees of freedom.  This effective field has the property
that a derivative acting on $h(x)$ yields the residual momentum $k^\mu$
rather than the full momentum $p_b^\mu$.  An expansion in terms of ${\rm
D}_\mu/m_b$ then becomes sensible. Expanding in powers of $1/m_b$, we
may invert \hintro\ to find
\eqn\bofh{
    \b(x)=\e^{-\im m_bv\cdot x}
    \left[ 1+{\im\Dslash/2m_b}+\cdots \right] h(x)\,.}
Inserting this into the usual QCD lagrangian $\ol\b\,\im\Dslash\b$, we
find the effective lagrangian for HQET \hqet,
\eqn\lhqet{
    \L=\ol h v\cdot\id h + \delta\L\,}
where the correction terms \ref\oneoverm{A.F. Falk, B. Grinstein and M.
Luke, {\sl Nucl. Phys.} B357 (1991) 185; E. Eichten and B. Hill, {\sl
Phys. Lett.} B243 (1990) 427}
\eqn\deltal{
    \delta\L={1\over2m_b}\,\ol h(\id)^2 h
    -{1\over2m_b}\,Z_1(\mu)\,\ol h(v\cdot\id)^2 h
    +{1\over2m_b}\,Z_2(\mu)\,\ol hs^{\mu\nu}G_{\mu\nu} h
    +O(1/4m_b^2)}
are treated as perturbations to the $m_b\to\infty$ limit.  Here the
gluon field strength is defined by $G_{\mu\nu}=[\id_\mu,\id_\nu]$,
and $s^{\mu\nu}=-\frac\im2\sigma^{\mu\nu}$.  The renormalisation
constants are given by
\eqn\renorm{\eqalign{
    Z_1(\mu)&=3\left({\alpha_s(m_b)\over\alpha_s(\mu)}\right)
    ^{8/25}-2\,,\cr
    Z_2(\mu)&=\left({\alpha_s(m_b)\over\alpha_s(\mu)}\right)
    ^{9/25}\cr}}
above the charm threshold.

Because the operator product expansion \opprod\ is an expansion in
$\D_\mu/m_b$, we must express the operators $\O_n$ in terms of the
HQET field $h(x)$ rather than the full fields $\b(x)$.  However, as we
shall see, it turns out to be convenient to leave the {\it leading\/}
operator in terms of $\b(x)$, and to expand the rest in $h(x)$.
The operators $\O_n$ which appear in the expansion \opprod\ then take
the form
\eqn\onform{\eqalign{
    \O_0&=\ol\b\,\Gamma\,\b\,,\cr
    \O_1&=\ol h\,\Gamma\,\id_\mu h\,,\cr
    \O_2&=\ol h\,\Gamma\,\id_\mu\id_\nu h\,,\cr}}
and so forth.  In each case, $\Gamma$ denotes an arbitrary Dirac
structure, in which we also absorb all dependence on the external
momentum $q$, as well as on any other variables.  We will keep
operators in the expansion with up to two derivatives.

We now turn to the evaluation of the forward matrix elements of the
operators $\O_n$ between $B$ meson states. At leading order, we need
matrix elements of the form
\eqn\lowmat{
    \langle B|\,\ol\b\,\Gamma\,\b\,|B\rangle\,,}
which is nonzero only for $\Gamma=1$ or $\Gamma=\gamma^\mu$.
In the second case, the conservation of the b-number current in QCD
yields the matrix element normalised absolutely,
\eqn\conserve{
    \langle B|\,\ol\b\gamma^\mu\b\,|B\rangle=2P_B^\mu\,.}
This, of course, is why we left $\O_0$ in terms of the field $\b(x)$ in
eq.~\onform. As for the scalar current, it may be rewritten in terms of
the vector current plus higher dimension operators of the form of
$\O_2$ \russians,
\eqn\scalar{
    \ol\b\,\b=v_\mu\ol\b\gamma^\mu\b+
    {1\over2m_b^2}\,\ol h\left[(\id)^2-(v\cdot\id)^2
    +s^{\mu\nu}G_{\mu\nu}\right] h+\ldots\,.}
This identity may easily be proven by using eq.~\bofh\ to expand both
sides in terms of the effective field $h$.  It is only meaningful when
the four-velocity $v^\mu$ of the b field is fixed.  The correction term
in eq.~\scalar\ may be absorbed into $\O_2$.  Hence, the leading term
in the expansion of $T(q)$ may be evaluated unambiguously, using
eq.~\conserve.  In fact, the leading term is precisely the free quark
decay model result, which becomes exact in the limit $m_b\to\infty$
\chayetal. The subleading operators $\O_n$ in the operator product
expansion \opprod\ will provide systematically the corrections for
finite b quark mass.

The evaluation of the matrix elements of the higher dimension operators
$\O_1$ and $\O_2$ involves the equation of motion of the effective
theory \ref\theorem{M. Luke, {\sl Phys. Lett.} B252 (1990) 447}.  This
is given by the lowest order lagrangian,
\eqn\eom{
    v\cdot\id h=0\,.}
Since the external states are characterized only by their four-velocity
$v^\mu$, Lorentz invariance severely restricts the forward matrix
elements of operators of the form \onform.  For the operator $\O_1$ of
dimension four, we find
\eqn\dimfour{
    \langle M|\,\O_1\,|M\rangle=
    \langle M|\,\ol h\,\Gamma\,\id_\mu h\,|M\rangle=
    \langle M|\,\ol h\,\Gamma v_\mu v\cdot\id h\,|M\rangle\,.}
However, this is now the matrix element of an operator which vanishes
by the equation of motion \eom.  Politzer \ref\hdavid{H.D. Politzer,
{\sl Nucl. Phys.} B172 (1980) 349} has shown that all such
matrix elements vanish identically; his proof is outlined in the
Appendix.  Since $\O_1$ is the only possible source of corrections of
order $1/m_b$ to the lowest order result, we see that the leading
corrections to the free quark decay model are actually of second order
in the heavy quark expansion.  As first pointed out by Chay {\it et
al.}, this is a most surprising result, since {\it exclusive\/} decay
modes all presumably receive corrections already at order $1/m_b$.
Somehow these individual contributions must cancel in the inclusive
rate.

The dimension five operators do give nonvanishing contributions, of
order $1/m_b^2$.   However, their forward matrix elements have a very
simple parameterisation \ref\falkneu{A.F. Falk and M. Neubert, {\sl
Phys. Rev.} D47 (1993) 2965; {\sl Phys. Rev.} D47 (1993) 2982}. The
symmetries of the effective theory may be used to write the matrix
element as an ordinary Dirac trace,
\eqn\dimfive{
    \langle M|\,\O_2\,|N\rangle=
    \langle M|\,\ol h\,\Gamma\,\id_\mu\id_\nu h\,|M\rangle=
    M_B\,{\rm Tr}\big\{\Gamma P_+\psi_{\mu\nu}P_+\big\}\,,}
where
\eqn\psidef{
    \psi_{\mu\nu}=\frac13\lambda_1(g_{\mu\nu}-v_\mu v_\nu)
    +\frac12\lambda_2\im\sigma_{\mu\nu}\,.}
The mass parameters $\lambda_1$ and $\lambda_2$ are defined in terms of
certain expectation values in the effective theory,
\eqn\lamdefs{\eqalign{
    &\langle M^{(*)}|\,\ol h(\id)^2 h\,|M^{(*)}\rangle
    =2M_B\lambda_1\,,\cr
    &\langle M^{(*)}|\,\ol hs^{\mu\nu}G_{\mu\nu} h\,|M^{(*)}\rangle
    =2M_Bd_{M^{(*)}}\lambda_2(\mu)\,,\cr}}
where $d_M=3$ and $d_{M^*}=-1$. The $\mu$-dependence of $\lambda_2$
cancels that of the renormalisation constant $Z_2(\mu)$ \renorm.
We note that $Z_2(m_b)=1$; hence from this point on we will drop
it and by $\lambda_2$ mean $\lambda_2(m_b)$.

The role which these parameters play in the effective theory is revealed
when one expands the masses of the heavy pseudoscalar and vector mesons
in powers of $1/m_b$:
\eqn\massexpand{\eqalign{
    M_B&=m_b+\ol\Lambda-{1\over2m_b}(\lambda_1+3\lambda_2)
    +\ldots\,,\cr
    M_{B^*}&=m_b+\ol\Lambda-{1\over2m_b}(\lambda_1-\lambda_2)
    +\ldots\,.\cr}}
In this expansion, the term $\ol\Lambda$ represents the energy of the
light degrees of freedom in the meson.  We see that $\lambda_1$ and
$\lambda_2$ are higher order effects of the finite b quark mass;
$\lambda_1$ is essentially a ``Fermi motion'' effect, while
$\lambda_2$, the leading spin symmetry-violating correction, arises
from the hyperfine chromomagnetic interactions.  From \massexpand\ we
have the well-known relation
\eqn\lamtwoexpt{\lambda_2={m_b\over 8}(M_{B^*}-M_B)\approx
    {1\over 4}(M_{B^*}^2-M_B^2)=0.12\gev^2\,,}
where we are neglecting higher-order corrections in $1/m_b$.
There have been attempts to extract both
$\lambda_1$ and $\lambda_2$ from QCD sum rules by computing the
matrix elements \lamdefs, with the results
\ref\mattias{M. Neubert, {\sl Phys. Rev.} D46 (1992) 1076}
\eqn\lamresults{\eqalign{
    &0\le\lambda_1\le1\gev^2\,,\cr
    &\lambda_2(1\gev)=0.12\pm0.02\gev^2\,.\cr}}
The parameter $\lambda_2$ is much better determined in this approach and
agrees nicely with the experimental $B$--$B^*$ mass splitting \lamtwoexpt.

Finally, there is one other source of corrections of order $1/m_b^2$,
namely time-ordered products of $\O_1$ with the correction $\delta\L$
to the effective lagrangian \lhqet.  As discussed in the Appendix,
these arise because the states $|M\rangle$ in the effective theory
differ at order $1/m_b$ from those $|B\rangle$ of QCD.  The difference
is compensated for at each order by computing matrix elements of the
form \theorem\falkneu
\eqn\timeord{\eqalign{
    &\im\int\d x\,\langle M|\,T\big\{\ol h\,\Gamma\,
    h\,,\delta\L(x)\big\}|M\rangle\,,\cr
    &\im\int\d x\,\langle M|\,T\big\{\ol h\,\Gamma\,\id_\mu h\,,
    \delta\L(x)\big\}|M\rangle\,,\cr}}
and so on.  We did not encounter the time-ordered product $\delta\L$
with the dimension three operator $\O_0$ above, because we were able to
compute its matrix element \conserve\ directly in full QCD. For the
operators of dimension four, since we are working in the effective
theory, we must evaluate eq.~\timeord. First, we use the fact that the
external states depend only on the four-velocity $v^\mu$ to write the
analogue of eq.~\dimfour:
\eqn\tord{
    \im\int\d x\,\langle M|\,T\big\{\ol h\,\Gamma\,\id_\mu h,\,
    \delta\L(x)\,|M\rangle=\im\int\d x\,\langle M|\,T\big\{
    \ol h\,\Gamma\, v_\mu v\cdot\id h,\,\delta\L(x)\big\}|M\rangle\,.}
We may now apply the identity derived in the Appendix, which exploits
the fact that the operator which appears on the right-hand side of
eq.~\tord\ vanishes by the equation of motion of the effective
theory.  We then obtain the matrix element of a local current,
\eqn\tordii{
    \im\int\d x\,\langle M|\,T\big\{\ol h\,\Gamma\,\id_\mu h,\,
    \delta\L(x)\big\}|M\rangle=-v_\mu\cdot{1\over2m_b}\langle M|\,
    \ol h\Gamma P_+F(\D)h\,|M\rangle\,,}
where $F(\D)=(\id)^2+s^{\mu\nu}G_{\mu\nu}$.  This matrix element may
then be evaluated in terms of $\lambda_1$ and $\lambda_2$, using
eq.~\dimfive.

The result of this long and involved procedure is an expression for the
nonperturbative hadronic quantity $T(q)$, of the form
\eqn\texpand{
    T(q)=T_0(q^2,v\cdot q)+{1\over4m_b^2}T_2(q^2,v\cdot q)+\ldots\,.}
The expansion is a series in $1/m_b$ and $\alpha_s(m_b)$, and the
ellipses in \texpand\ denote higher order terms in both small
parameters.  (The radiative corrections to $T_0$ have been computed
previously \ref\ali{A. Ali and C. Greub, {\sl Z. Phys.} C49 (1991)
431}; we will not include those to $T_2$.)  We now take the imaginary
part of $T(q)$ to recover $W(q)$, multiply by the perturbative part of
the matrix element which couples to $P(\phi)$, and compute the
inclusive differential width $\d\Gamma$.  This may then be integrated
to give the total width, $\Gamma$, or other smoothly weighted
distributions.  As we have mentioned, the contribution of $T_0$  will
be precisely that of the free quark decay model \chayetal. The leading
corrections to the $m_b\to\infty$ limit are of relative order $1/m_b^2$
and encoded in $T_2$; they are expressible entirely in terms of the
mass parameters $\lambda_1$ and $\lambda_2$.  We will now apply this
procedure, and compute these corrections, for two interesting examples.

\newsec{Application to Rare $B$ Decays}

We will consider inclusive decays of the form $\bsg$ and $\bsee$, where
$\ell=\e$ or $\mu$ is a light lepton. They are governed by the
effective Hamiltonian density
\eqn\hamdens{\CH_{\rm eff}=-{4 G_F\over\sqrt{2}}\km
    \sum_j c_j(\mu) O_j(\mu)\,,}
where the sum is over the truncated set of local operators
\eqn\bsgops{\eqalign{
    &O_1=\sbar_\alpha\gamma^\mu P_L\b_\alpha\,\bar\c_\beta
    \gamma_\mu P_L\c_\beta\cr
    &O_2=\sbar_\alpha\gamma^\mu P_L\b_\beta\,\bar\c_\beta
    \gamma_\mu P_L\c_\alpha\cr
    &O_7={e\over 16\pi^2}m_b\,\sbar_\alpha\sigma^{\mu\nu}P_R
    \b_\alpha\,F_{\mu\nu}\cr
    &O_8={e^2\over 16\pi^2}\,\sbar_\alpha\gamma^\mu P_L\b_\alpha
    \ol\e\gamma_\mu\e\cr
    &O_9={e^2\over 16\pi^2}\,\sbar_\alpha\gamma^\mu P_L\b_\alpha
    \ol\e\gamma_\mu\gamma_5\e\,,}}
Here $P_L=\gamleft$ and $P_R=\gamright$ are helicity projection
operators, $F_{\mu\nu}$ is the photon field strength and $\alpha$,
$\beta$ are colour indices.  We have included in $O_8$ and $O_9$ only
the coupling to the electron current; the coupling to the muon is
analogous.  There are also additional operators, such as
$\sbar_\alpha\gamma^\mu P_L b_\alpha\,[\bar\u_\beta\gamma_\mu P_L
\u_\beta+\dots+ \bbar_\beta\gamma_\mu P_L\b_\beta]$, which contribute
to these decays, but their coefficients are small and we shall neglect
them. The coefficients $c_j(m_b)$ have been calculated in leading
logarithmic approximation, both in the standard model and in certain
minimal extensions, and are presented in refs.~\ref\gswa{B. Grinstein,
R.P Springer and M.B. Wise, {\sl Phys. Lett.} B202 (1988) 138\semi {\sl
Nucl. Phys.} B339 (1990) 269}\nref\gosn{R. Grigjanis, P.J. O'Donnell,
M. Sutherland and H. Navelet, {\sl Phys. Lett.} B213 (1988) 355; {\sl
Phys. Lett.} B223 (1989) 239; {\sl Phys. Lett.} B237 (1990) 252; {\sl
Phys. Rev.} D42 (1990) 245}\nref\ccrv{G. Cella, G. Curci, G. Ricciardi
and A. Vicere, {\sl Phys. Lett.} B248 (1990) 181}\nref\crv{G. Cella, G.
Ricciardi and A. Vicere, {\sl Phys. Lett.} B258 (1991) 212}\nref\mis{M.
Misiak, {\sl Phys. Lett.} B269 (1991) 161; {\sl Nucl. Phys.} B393
(1993) 23}\nref\gsw{B. Grinstein, M.J. Savage and M.B. Wise, {\sl Nucl.
Phys.} B319 (1989) 271}--\ref\cfmrs{M. Ciuchini, E. Franco, G.
Martinelli, L. Reina and L. Silvestrini, LPTENS 93/28, ROME 93/958 and
ULB-TH 93/09 (1993)}

We now apply the procedure of the previous section to compute the rates
for inclusive decays mediated by the operators \bsgops.  The first step
is to construct the operator product expansion \opprod, which takes the
form
\eqn\fullglory{\eqalign{
    T\big\{\bbar\gone\s,\sbar\gtwo\b\big\}&\mathrel{\mathop=^{\rm OPE}}
    {1\over m_b}\left[\CO_0 +{1\over 2m_b}\CO_1
    +{1\over 4m_b^2} \CO_2 +\dots\right].}}
For now, we will allow $\gone$ and $\gtwo$ to be arbitrary Dirac
matrices. To compute the terms in this series, we must expand the
diagrams in \ope\ in powers of $1/m_b$.  Fixing the four-velocity of
the external b quark to be $v^\mu$, we may expand its momentum as
$p_b^\mu=m_b v^\mu+k^\mu$.  Then the graph in \ope (a) gives
\eqn\workthrough{\eqalign{
    \im\CM&=-\im\bar u_b\,{\gone\,(m_b \vsl-\qsl+\ksl+\ms)\,\gtwo \over
    (m_b v-q+k)^2-\ms^2+\im\epsilon}\,u_b\cr
    &=-{\im\over m_b x}\,\bar
    u_b\gone\,(\vsl-\qhsl+\mshat)\,\gtwo u_b\cr
    &\qquad-{\im\over m_b^2}\,\bar u_b\({1\over x}\gone\ksl\gtwo
    -{2\over x^2}\gone\,(\vsl-\qhsl+\mshat)\gtwo\,
    (v-\qhat)^\alpha k_\alpha\)u_b\cr&\qquad+O\({1/m_b^3}\)}}
where $\qhat=q/m_b$, $\mshat=\ms/m_b$, and
\eqn\poleloc{
    x=1-2 \vdotq+\qhat^2-\hat m_s^2+\im\epsilon}
contains the pole corresponding to an on-shell strange quark, near the
end of the physical cut.  The spinor $u_b$ which appears is the
ordinary on-shell b quark spinor of QCD.  From the matrix element
\workthrough, we may deduce the first two terms in the operator product
expansion \fullglory:
\eqn\leadingo{\eqalign{
    &\O_0={1\over x}\,\bbar\,\gone(\vsl-\qhsl+\mshat)\gtwo\,\b\,,\cr
    &\O_1={2\over x}\,\ol h\,\gone\gamma^\alpha\gtwo\id_\alpha h
    -{4\over x^2}(v-\qhat)^\alpha\,\ol h\,\gone(\vsl-\qhsl+\mshat)
    \gtwo\,\id_\alpha h\,.}}
Note that while we have left the leading operator in terms of the
four-component fields $\b(x)$, we have expanded $\O_1$ in terms of the
two-component effective fields $h(x)$.  The reason for this choice was
discussed in the previous section.

To obtain $\O_2$ we must also expand the one-gluon graph in \ope (b),
in order to identify the contribution from the gluon field strength
$G_{\mu\nu}=[\id_\mu,\id_\nu]$.  Additional contributions to $\O_2$
arise when the full QCD fields $\b(x)$ in $\O_1$are replaced by the
effective theory fields $h(x)$ via the relation \bofh.  Equivalently,
one may expand the spinors $u_b$ in the matrix element \workthrough\ in
terms of the two-component spinors $u_h$ of HQET,
\eqn\spinors{u_b=\left[1+{\ksl\over 2m_b}+O\({1\over
    m_b^2}\)\right]u_h\,,}
and check that the result may be made covariant. Finally, there will be
corrections at order $1/m_b^2$ if the leading operator $\O_0$ contains
a scalar current $\bbar\b$, because of the expansion \scalar. These are
still contained in $\O_0$ and are {\it not\/} included in $\O_2$. A
straightforward calculation then yields
\eqn\secondo{\eqalign{
    \O_2=&{16\over x^3}(v-\qhat)^\alpha(v-\qhat)^\beta\,\ol h\,\gone
    (\vsl-\qhsl+\mshat)\gtwo\,\id_\alpha\id_\beta h\cr
    &-{4\over x^2}\,\ol h\,\gone(\vsl-\qhsl+\mshat)\gtwo
    (\id)^2h\cr
    &-{4\over x^2}(v-\qhat)^\beta\,\ol h\,\gone\gamma^\alpha
    \gtwo\,(\id_\alpha\id_\beta+\id_\beta\id_\alpha)h\cr
    &+{2\over x^2}\hat m_s\,\ol h\,\gone\,\im\sigma_{\alpha\beta}
    \gtwo G^{\alpha\beta}h\cr
    &-{2\over x^2}\im\epsilon^{\mu\lambda\alpha\beta}
    (v-\qhat)_\lambda\,\ol h\,\gone\gamma_\mu\gamma_5\gtwo
    G_{\alpha\beta}h\cr
    &+{2\over x}\,\ol h\(\gamma^\beta\gone\gamma^\alpha\gtwo
    +\gone\gamma^\beta\gtwo\gamma^\alpha\)\id_\beta\id_\alpha h\cr
    &-{4\over x^2}(v-\qhat)^\alpha\,\ol h\,\gamma^\beta\gone
    (\vsl-\qhsl+\mshat)\gtwo\,\id_\beta\id_\alpha h\cr
    &-{4\over x^2}(v-\qhat)^\alpha\,\ol h\,\gone
    (\vsl-\qhsl+\mshat)\gtwo\gamma^\beta\,\id_\alpha\id_\beta h\,.}}
To continue any further, we must specify the Dirac structures
$\Gamma_1$ and $\Gamma_2$.

\subsec{$\bsg$}

For the transition $\bsg$, only the operator $O_7$ from \bsgops\
contributes and the operator product expansion simplifies considerably.
We now contract the terms in the time ordered product \fullglory\ with
the external photon fields and take the matrix element between $B$
mesons to construct the hadronic object $T(q)$ defined in eq.~\tdef.
Because the decay is to an on-shell photon, $q^2=0$ is fixed, and
$T(q)$ becomes a function only of the scaled photon energy
$\vdotq=E_\gamma/m_b$. Including the matrix element of the photon
field, we find
\eqn\tbsg{\eqalign{
    \tilde T(\vdotq)&\equiv\im^2\langle B|\,
    T\big\{\bbar\sigma^{\mu\nu}\s,\sbar\sigma^{\rho\tau}\b\big\}
    |B\rangle \cdot\sum_{\epsilon=1,2}\langle\gamma(\qhat,\epsilon)|
    \,F_{\mu\nu}F_{\rho\sigma}\,|\gamma(\qhat,\epsilon)\rangle\cr
    &=-16 M_B m_b (\vdotq)^2
    \left[{1\over x}-{\lamone\over 2 m_b^2}\({5-6\vdotq\over3x^3}\)
    +{\lamtwo\over 2m_b^2}\cdot{3\over x^2}\right].}}
The sum is over the transverse polarizations of the photon, and there
is a factor of $\im$ from each insertion of the effective Hamiltonian
\hamdens. We neglect contributions of order $\hat m_s^2$. Note that we
distinguish between $m_b$, the bottom {\it quark\/} mass which arises
in the operator product expansion, and $M_B$, the $B$ {\it meson} mass
which arises from the relativistic normalisation of the states (and
therefore drops out of the final expression). The inclusive rate for
$B\rightarrow X_s\gamma$ is then given by
\eqn\gambsg{\eqalign{\Gamma_{B\rightarrow X_s\gamma}
    &={\alpha G_F^2\over 8 \pi^3}{m_b^2\over M_B} |\km|^2 |c_7(m_b)|^2
    \,{\rm Im}\,\int\,\dtk\,\tilde T(\vdotq)\cr
    &={\alpha G_F^2\over 32\pi^5}{m_b^4\over M_B} |\km|^2
    |c_7(m_b)|^2 \oint\, z\,\tilde T(z)\,\d z\,,}}
where $z=\vdotq$, and the contour integral is taken around
the pole at $x=0$.  It is straightforward to evaluate
this integral, and we find
\eqn\widbsg{\Gamma_{B\rightarrow X_s\gamma}
    ={\alpha G_F^2\over 16\pi^4}
    m_b^5 |\km|^2 |c_7(m_b)|^2\left[1+{1\over 2m_b^2}
    \(\lamone-9\lamtwo\)\right].}
This expression for the total rate agrees with the result of
ref.~\russians.  The first term is just what one would obtain in the
free quark decay model.

We may consider using the same method to compute certain features of
the photon energy spectrum.  Of course, the precise shape of the
spectrum is not available to us, in particular its behaviour near the
endpoint of maximum $E_\gamma$.  This is true at {\it any\/} order in
the $1/m_b$ expansion, because in this region the strange quark
approaches its mass shell and the operator product expansion breaks
down. For example, the fact that the true endpoint of the photon energy
spectrum is found not at $E_\gamma=m_b/2$ but rather at
$(M_B^2-M_K^2)/2M_B$ is entirely unavailable to us in this formalism.

It is instructive, however, to generalise eq.~\gambsg\ to calculate the
nonperturbative contributions to the moments of the energy spectrum.
For example, the deviation of the average photon energy from that of
the free quark decay model is
\eqn\moments{\eqalign{
    \langle E_\gamma \rangle=&{1\over\Gamma_{\bsg}}
    {\alpha G_F^2\over32\pi^5}{m_b^4\over M_B}
    |\km|^2 |c_7(m_b)|^2 \oint\, (m_b z)\,z\,\tilde T(z)\,\ dz\cr
    =&{m_b\over
    2}\(1-{\lamone+3\lamtwo\over 2 m_b^2}\). }}
This result has an interesting structure, if we compare it to the
$1/m_b$ expansion of the $B$ meson mass \massexpand.  Na\"\i vely, we
might have expected $\langle E_\gamma \rangle$ to be shifted from the
free quark value $m_b/2$ to half the physical meson mass $M_B/2$.
However, that is not what we find; only the order $1/m_b^2$ terms
contribute.  The reason is that the correction $\ol\Lambda$ to $M_B$ in
eq.~\massexpand\ is the contribution to the meson mass of the light
antiquark and the other light degrees of freedom, which in this
formalism are mere spectators to the decay.  Since they are present in
the final state $X_s$ as well as in the initial state, they do not
represent additional energy available to the photon.  By contrast, the
higher-order mass corrections proportional to $\lambda_1$ and
$\lambda_2$ arise from terms in the effective Hamiltonian of the b
quark, representing its ``Fermi motion'' and its chromomagnetic
interaction with the soft hadronic surroundings.  These bound state
shifts in the b quark energy are then reflected in the average photon
energy $\langle E_\gamma \rangle$.

We could also generalise eq.~\moments\ to higher moments of the photon
spectrum.  However, there arises an additional complication if we
insist that the moments we compute be experimentally meaningful
quantities.  This is because they are constructed by convolving a power
of the photon energy with the measured energy spectrum, but this
spectrum is only related to our computation once our result \tbsg\ has
been smeared over typical hadronic scales. That is, $\tilde T(z)$
should be replaced by the smoothed quantity
\eqn\tsmeared{
    \tilde T_f(z)=\int\d z'\, f(z-z')\,\tilde T(z')\,,}
where $f(x)$ is some smearing function of width $\delta=\Delta E/m_b$.
If we take $f(x)$ to be a Gaussian distribution,
$f(x)=\exp(-x^2/\delta^2)/\sqrt{\pi\delta^2}$, we can calculate the
moments analytically.  Keeping terms of order $\delta^2$, for the
$n$'th moment we find
\eqn\moments{\eqalign{
    \Gamma^{(n)}=&{\alpha G_F^2\over32\pi^5}{m_b^4\over M_B}
    |\km|^2 |c_7(m_b)|^2 \oint\, (m_b z)^n\,z\,\tilde T_f(z)\,\ dz\cr
    =&\({m_b\over2}\)^n\left[
    1+2n(n-1){(\Delta E)^2\over2m_b^2}
    -{n(n+2)\over3}{\lambda_1\over2m_b^2}-3n{\lambda_2\over2m_b^2}
    \right]\Gamma_{\bsg}\,.}}
The total rate ($n=0$) and average energy ($n=1$) which we have already
presented are unaffected by this procedure, but the same is not true
for the moments with $n\ge2$.  Note that the effect of the smearing is
proportional to $\delta^2$ rather than $\delta$, and so is formally of
the same order as the nonperturbative corrections we have been
considering. However, in order for our inclusive predictions to be
meaningful, the resolution $\Delta E$ with which the photon energy
spectrum is measured actually must be much greater than $\lqcd$, so
that many exclusive states are always summed over.  Hence, it is in
fact this resolution, rather than the nonperturbative effects,
which will dominate the corrections to the moments $\Gamma^{(n)}$.  In
addition, real gluon emission will broaden the energy spectrum over the
entire allowed phase space $0<E_\gamma<m_b/2$, which will affect
substantially the shape of the experimentally measured spectrum \ali.

\subsec{$\bsee$}

The transition $\bsee$ receives contributions from the complete set of
operators in $\bsgops$.  In particular, unlike the decay $B\rightarrow
X_s\gamma$, the four-quark operators $O_1$ and $O_2$ have non-vanishing
matrix elements. In order for our treatment of the four-quark operators
to be valid, it is crucial that the invariant mass of the lepton pair
not be near any resonances in the charm system such as the $\psi$, so
that strong final state interaction corrections will be small. In this
case we can treat the contributions from $O_1$ and $O_2$ as effectively
local on the scale of hadronic interactions.

It is convenient to separate the total rate for $\bsee$ into two terms,
corresponding to the decay to left- and right-handed leptons
\eqn\matee{
    \d\Gamma_\bsee \propto
    W^{\mu\nu}_L(\qhat^2,\vdotq)L^L_{\mu\nu}
    \ +\ W^{\mu\nu}_R(\qhat^2,\vdotq) L^R_{\mu\nu}\,,}
where the lepton tensors are given by
\eqn\lep{
    L^{L (R)\mu\nu} = p_+^\mu p_-^\nu + p_+^\nu p_-^\mu
    - g^{\mu\nu}p_+\cdot p_
    -\pm\im\epsilon^{\mu\nu\sigma\rho}p_{+\sigma}p_{-\rho} \,.}
Here $p_+$ and $p_-$ are respectively the four-momenta of the $\ell^+$
and $\ell^-$.  Since we are restricting ourselves to $\ell=\e$ or
$\mu$, we neglect the masses of the leptons.

The two Lorentz structures which arise from the effective Hamiltonian
\bsgops\ are $\gamma_\mu(1-\gamma_5)$ and
$\sigma_{\mu\nu}(1+\gamma_5)\qhat_\nu$.  Hence it is convenient to write
\eqn\gam{\eqalign{
    \Gamma^{L(R)}_2 = &\gamleft\gamma_\mu
    \left[A^{L(R)} - B^{L(R)}\qhsl/\shat\right],\cr
    \Gamma_1^{L (R)} = &\Gamma_2^{{L (R)}\dagger}\,,}}
where $\shat=\qhat^2$. The $A$'s and $B$'s are then combinations of the
coefficients $c_1,\ldots,c_9$:
\eqn\coeffs{\eqalign{
    A^L &= c_8(m_b) - c_9(m_b)
    + \left[3c_1(m_b)+c_2(m_b)\right] g(m_c/m_b,\hat s) \,,\cr
    A^R &= c_8(m_b) + c_9(m_b)
    + \left[3c_1(m_b)+c_2(m_b)\right] g(m_c/m_b,\hat s) \,,\cr
    B^L &= B^R = -2c_7(m_b)\,,}}
where the function $g(m_c/m_b,\hat s)$ multiplying $c_1$ and $c_2$
arises from taking the one-loop matrix elements of $O_1$ and $O_2$ and
has the form
\eqn\charm{\eqalign{
    g(z,\hat s) &=
    -{4\over 9}\log z^2 + {8\over 27} + {16\over 9}{z^2\over\hat s} -
    {2\over 9}\sqrt{1-{4z^2\over \hat s}}\left(2+{4z^2\over\hat
    s}\right)\cr
    &\qquad\qquad\qquad\qquad\times\log\left({\sqrt{1-4z^2/\shat}
    +1+\im\epsilon\over\sqrt{1-4z^2/\shat}-1+\im\epsilon}\right).}}
Integrating over the lepton phase space, the total decay rate is given
by
\eqn\bseepp{\eqalign{
    \Gamma_\bsee&={1\over 2M_B}\int\,
    {\d^3 p_+\over (2\pi)^3 2 E_+}\, {\d^3 p_-\over (2\pi)^3 2 E_-}
    \(W^{\mu\nu}_L L_{\mu\nu}^L+W^{\mu\nu}_R L_{\mu\nu}^R\)\cr
    &={m_b^2\over 128\pi^4M_B}\int\,
    \d\shat\,\d\hat E_1\,{\rm Im} \oint \d\vdotq
    \(T^{\mu\nu}_L L_{\mu\nu}^L+T^{\mu\nu}_R L_{\mu\nu}^R\).}}
We must next perform the contour integral in the $\vdotq$ plane and
then the $\hat E_1$ integral to obtain the differential decay width.
Since the calculation is quite tedious and the intermediate expressions
extremely lengthy, we present only the final result:
\eqn\rateee{\eqalign{
    {\d\Gamma_\bsee\over\d\shat}
    & = {G_F^2\alpha^2\over 256\pi^5}m_b^5|\km|^2 (1-\shat)\cr
    &\quad\times\sum_{i=L,R}\bigg\{
    \coeff16 (1-\shat)(1+2\shat)|A^i|^2 + \coeff16 (1-\shat)
    (1+2/\shat)|B^i|^2\cr
    &\qquad\qquad\qquad-(1-\shat){\rm Re}\,(B^{*i}A^i)\cr
    &\qquad\qquad+{\lambda_1\over 2m_b^2}
    \big[\(-\coeff13\shatsq +
    \coeff12\shat + \coeff56\)|A^i|^2
    -\coeff16(1+\shat)|B^i|^2\cr
    &\qquad\qquad\qquad\qquad+\(\shat
    - \coeff53\){\rm Re}\,(B^{*i}A^i)\big]\cr
    &\qquad\qquad+{\lambda_2\over 2m_b^2} \big[
    \(-5\shatsq + \coeff{15}2\shat + \coeff12\)|A^i|^2
    -\coeff52(1+\shat)|B^i|^2\cr
    &\qquad\qquad\qquad\qquad+(7\shat-5){\rm Re}\,(B^{*i}A^i)
    \big]\bigg\}.\cr}}
The summation is over the two chirality states of the leptons.

The leading term in eq.~\rateee\ reproduces the free quark decay model
result obtained in refs.~\gosn\gsw, while the subsequent terms are the
leading non-perturbative contributions to the decay rate. It is
interesting to note that unlike the parton level result, which has a
characteristic $1/\shat$ behaviour at small $\shat$ from the one-photon
intermediate state, the non-perturbative corrections approach a finite
constant value as $\shat\to 0$. The differential spectrum for the
invariant mass of the lepton pair is plotted in \fig\bseespec{Invariant
mass spectrum for $\bsee$. The solid line corresponds to the
parton model, while the dashed line corresponds to
$\lambda_1=0.5\gev^2$ and $\lambda_2=0.12\gev^2$.  The cusp at $\shat =
(2m_c/m_b)^2$ corresponds to the charm threshold.  Near this point our
estimate of the nonperturbative corrections is not valid due to
resonance effects.}.  We have chosen a top quark mass of $m_t=150\gev$,
along with $m_b = 4.5\gev$, $\alpha_s(m_W) = 0.12$ and $\alpha_s(m_b) =
0.21$, to generate the spectrum. The free quark decay model result
($\lambda_1=\lambda_2=0$) is presented along with the spectrum for
$\lambda_1=0.5\gev^2$ and $\lambda_2=0.12\gev^2$. We have normalised
the width for this decay to that for semileptonic $B$ decay (which
includes the nonperturbative corrections given in
ref.~\ref\semileptonic{I.I. Bigi, M. Shifman, N.G. Uraltsev and A.
Vainshtein, {\sl Phys. Rev. Lett.} 71 (1993) 496; B. Blok, L. Koyrakh,
M. Shifman and A.I. Vainshtein, NSF-ITP-93-68 (1993); A. Manohar and
M.B. Wise, UCSD/PTH 93-14, hep-ph/9308246 (1993); T. Mannel, IKDA
93/26, hep-ph/9308262 (1993)}). The modification to the $\bsee$ rate is
reasonably large and tends to enhance the overall rate for high mass
lepton pairs by order $10\%$.

\newsec{Summary and Conclusions}

Because of the necessary cuts to remove backgrounds, the full spectrum
from a decay such as $B\rightarrow X_s\gamma$ and $\bsee$ is not
available in an accelerator experiment.  It is therefore important to
understand well the shapes of these spectra if one is to relate the
observed branching fractions to fundamental parameters of the
electroweak theory. This is particularly true for the high photon
energy and high invariant lepton mass regions of the Dalitz plot.
Modifications to the simplest model, that of free quark decay, arise
from strong interactions that can be classified heuristically as
perturbative and non-perturbative corrections.

The perturbative corrections arising from gluon bremstrahlung and
one-loop effects for $\bsg$ have been computed previously \ali. It is
to the non-perturbative corrrections that we have addressed ourselves
in this paper. We have detailed the formalism for treating the
semi-hadronic inclusive decays of mesons containing a single heavy
quark.  Upon summing over all hadronic final states, one may express
the rate for a given process in terms of a time-ordered product of
quark bilinears. This time-ordered product is then expanded in a series
of local operators, the matrix elements of which either are known or
may be parameterised simply.  Heavy quark symmetries and the heavy
quark effective theory play a key role in the analysis.

We have applied these tools to the rare decays $B\rightarrow X_s\gamma$
and $B\rightarrow X_s\ell^+\ell^-$. The leading non-perturbative
corrections to the free quark decay model, of relative order $1/m_b^2$,
may be expressed entirely in terms of two low-energy parameters.  One
of these is determined from the splitting between the heavy
pseudoscalar and vector mesons; a model-dependent estimate of the other
comes from QCD sum rules. In addition to the total rates, we have
computed the correction to the average photon energy in $B\rightarrow
X_s\gamma$ and found the shift to be small. The correction to the
spectrum for $B\rightarrow X_sl^+l^-$ is larger and for high invariant
mass lepton pairs is at about the $10\%$ level.

Finally, we note that there has been considerable recent work in which
a similar formalism has been applied to semileptonic b decays
\semileptonic.

\bigskip\bigskip
\centerline{{\bf Acknowledgements}}
\bigskip
Much of this work was performed while in residence at the Aspen Center
for Physics, and it is a pleasure to thank them for their warm and
efficient hospitality.  We are grateful to A.~Manohar and M.~Wise for
helpful conversations.  This work was supported by the Texas National
Research Laboratory Commission under grants RGFY93--206 and FCFY9219.

\appendix{A}{Matrix Elements and the Equation of Motion}

In this appendix we derive an identity for the matrix element of a
time-ordered product of two operators, where one of the operators
vanishes by the equation of motion of the theory. This will be a
generalisation  of a proof by Politzer \hdavid\ that matrix elements of
single operators which vanish by the equation of motion themselves
vanish. We will derive our result within the context of the heavy quark
effective theory (HQET), because this is the application which we have
in mind, but with obvious modifications our result is completely
general.

We begin by recalling how such time-ordered products arise within HQET.
In this effective theory, the heavy quark part of the lagrangian takes
the form \hqet\oneoverm
\eqn\lagran{
    \L=\ol hv\cdot\id h + {1\over2m_b}\,\ol hF(\D)h + \ldots\,,}
where
\eqn\aofd{
    F(\D)=(\id)^2 - Z_1(\mu)(v\cdot\id)^2
    + Z_2(\mu)s^{\mu\nu}G_{\mu\nu}\,,}
and the ellipses denote terms of higher order in the $1/m_b$ expansion.
Here the gluon field strength is defined by
$G_{\mu\nu}=[\id_\mu,\id_\nu]$, and
$s^{\mu\nu}=-\frac\im2\sigma^{\mu\nu}$. The renormalisation  constants
$Z_1(\mu)$ and $Z_2(\mu)$ are given in Section 2. The equation of
motion in HQET is derived from the leading term in the lagrangian
\lagran, and is simply
\eqn\eomlead{
    v\cdot\id h=0\,.}
Instead of being included in the equation of motion, the corrections to
$\L$ in eq.~\lagran\ are treated as perturbations. They reappear in the
following way: because the states $|M\rangle$ of HQET are defined by
the truncated equation of motion \eomlead, they differ from those
$|B\rangle$ of full QCD.  The states $|M\rangle$ have the significant
advantage that, unlike $|B\rangle$, they are independent of the heavy
quark mass $m_b$, and so have simple transformations under the
spin-flavor symmetries of the effective theory.  The difference between
$|M\rangle$ and the physical states $|B\rangle$ is then compensated by
including in the matrix elements of effective operators additional
time-ordered products with the subleading terms in $\L$
\theorem\falkneu. That is, if we have an effective operator $\ol h'
C(\D) h$ whose matrix element we require between eigenstates
$|B\rangle$ of full QCD, then we must write
\eqn\time{\eqalign{
    \langle B(p')|\,\ol h' C(\D) h\,|B(p)\rangle
    &=\langle M(v')|\,\ol h' C(\D) h\,|M(v)\rangle\cr
    &\quad+ {1\over2m_b}\,\im\int\d x\,\langle M(v')|\,T\big\{\ol h'
    C({\rm D})h,\,\ol h F({\rm D}) h(x)\big\}|M(v)\rangle
    +\ldots\,.\cr}}
We have shown the expansion up to order $1/m_b$ explicity; the ellipses
denote denote terms of higher order which may be included if more
accuracy is needed.  We consider here the general case in which the
initial and final heavy quarks have different four-velocities.  The
field $h$ creates a heavy quark with velocity $v^\mu$, while $h'$
creates one with velocity $v'^\mu$.  There is a separate effective
lagrangian \lagran\ for each of these fields, but for simplicity we
will include the $1/m_b$ corrections only for the field $h$. The
time-ordered products in \time\ are new nonperturbative matrix elements
which must be evaluated if one wishes to use the effective theory
beyond leading order.

We will be concerned with a special case of eq.~\time, in which the
operator $\ol h' C(\D) h$ vanishes by the equation of motion \eomlead\
of the effective theory.\foot{We are grateful to A.~Manohar for
discussions of this point.  See also ref.~\semileptonic.} That is,
$C(\D)$ takes the particular form
\eqn\bform{
    C(\D)=A(\D)v\cdot\id\,,}
where $A(\D)$ may include an arbitrary Dirac structure.  Politzer
has shown that matrix elements of such operators, such as would appear
in the first term in eq.~\time, vanish \hdavid:
\eqn\firstbye{
    \langle M(v')|\,\ol h'A(\D)v\cdot\id h\,|M(v)\rangle=0\,.}
Note that it is the effective theory states $|M\rangle$ which appear
here. The purpose of this appendix is to generalise this argument to
prove a similar identity for the time-ordered product appearing in the
second term of \time, namely, that
\eqn\theidentity{\eqalign{
    \im\int\d x\,\langle M(v')|\,T\big\{\ol h'A(\D)v\cdot\id h,\,
    &\ol h F(\D)h(x)\big\}|M(v)\rangle\cr
    =&-\langle M(v')|\,\ol h'A(\D) P_+ F(\D) h\,|M(v)\rangle\,,\cr}}
where $P_+=\frac12(1+\vslash)$.  For the computation of this paper, we
will apply this identity in the case that $A(\D)$ actually contains no
derivatives, and at zero recoil; then the matrix element on the
right-hand side of \theidentity\ is of the simple form
\eqn\rhsform{
    \langle M|\,\ol h\,\Gamma\,\id_\mu\id_\nu h\,|M\rangle\,,}
and can be evaluated in terms of the constants $\lambda_1$ and
$\lambda_2$ as in Section 2.

Before proving the identity \theidentity, however, we note its relation
to the result we would obtain by taking a different approach. Instead
of introducing the states $|M\rangle$, which are eigenstates of the
lowest order effective lagrangian, we could chose to work always in
terms of the full states $|B\rangle$ of QCD.  This would be
undesireable, in that it would reintroduce the mass-dependence which it
is the goal of HQET to remove, thereby obscuring the spin-flavor
symmetries of the heavy quark limit.  However, if we do so, the
equation of motion is given by the full Lagrangian \lagran, taking the
form
\eqn\eombad{
    v\cdot\id h=-{1\over2m_b}P_+F(\D)h+\ldots\,.}
This equation of motion may be applied directly to matrix elements
between the states $|B\rangle$. We then find for the matrix element
\time\ the relation
\eqn\timeii{
    \langle B(p')|\,\ol h' C(\D) h\,|B(p)\rangle=
    -{1\over2m_b}\langle B(p')|\,\ol h'A(\D) P_+ F(\D)
    h\,|B(p)\rangle\,.}
Inserting eqs.~\firstbye\ and \theidentity\ into eq.~\time, and noting
that the states $|M\rangle$ and $|B\rangle$ differ only at order
$1/m_b$, we see that this is the same result that we find working
entirely within the effective theory.  The proof which we now present
may be seen as verifying the consistency of the effective theory
approach.  It is both an application and a generalisation of the proof
of Politzer.

We begin by using the LSZ reduction formula to write the desired
matrix
element in terms of a vacuum expectation value:
\eqn\lszapply{\eqalign{
    &\im\int\d x\,\langle M(v')|\,T\big\{\ol h'A(\D)v\cdot\id h(0),\,
     \ol h F(\D)h(x)\big\}|M(v)\rangle\cr
    &\qquad=\im\int\d x\int\d z\d z'\e^{\im k\cdot z}\e^{\im k'\cdot
     z'}v\cdot\id_z\, v'\cdot\id_{z'}\cr
    &\qquad\qquad\qquad\cdot\langle0|\,T\big\{\ol\q\Gamma'h'(z'),\,
     \ol h'A(\D)v\cdot\id h(0),\,\ol hF(\D)h(x),\,\ol h\Gamma\q(z)
     \big\}|0\rangle\,.\cr}}
Here the operators $\ol h\Gamma\q$ and $\ol\q\Gamma'h$ interpolate the
initial and final meson states, respectively.  (Of course, the proof is
valid for any external heavy hadrons, not just mesons.) Note that in
HQET, the one-particle poles are projected out by the differential
operator $v\cdot\id$ rather than by $[(\id)^2-M_B^2]$ as in full
QCD.

We now write the generating functional for Green's functions of this
theory:
\eqn\genfun{
    \exp(\im W)=\int[\d h][\d\ol h][\d h'][\d\ol h'][\d A_\mu]
    \exp\left\{\im\int\d y\,\left[\L_0+S_J+S_L+S_M+\ldots\right]
    \right\}\,.}
Here
\eqn\lzero{
    \L_0=\ol hv\cdot\id h+\ol h'v'\cdot\id h'}
is the lagrangian of the effective theory, and we have included
explicitly a variety of relevant source terms:
\eqn\sources{\eqalign{
    S_J&=J\,\ol h'A(\D)v\cdot\id h\,,\cr
    S_L&=L\,\ol hF(\D)h\,,\cr
    S_M&=K\,\ol h\Gamma\q + K'\,\ol\q\Gamma'h'\,.\cr}}
The ellipses denote sources for the fermions and gauge fields, and
gauge-fixing terms which will play no role in the analysis.  With these
definitions, then, we have
\eqn\firstdef{\eqalign{
    {\delta\over\delta\im K'(z')}{\delta\over\delta\im J(0)}
    &{\delta\over\delta\im L(x)}{\delta\over\delta\im K(z)}\,
    \exp(\im W)\bigg|_{{\rm sources}\,=0}\cr
    &=\langle0|\,T\big\{\ol\q\Gamma'h'(z'),\,
    \ol h'A(\D)v\cdot\id h(0),\,\ol hF(\D)h(x),\,\ol
    h\Gamma\q(z)\big\}|0\rangle\,.\cr}}
We now perform a shift of the integration variable $\ol h$,
\eqn\shift{
    \ol h=\ol h^* - J\,\ol h'A(\D)P_+\,,}
insert it into the generating functional \genfun, and drop terms of
order $J^2$.  We then obtain shifts in some of the expressions \lzero\
and \sources:
\eqn\shiftresult{\eqalign{
    \L_0&=\ol h^*v\cdot\id h - J\,\ol h'A(\D)v\cdot\id h
          + \ol h'v'\cdot\id h'\,,\cr
    S_J&=J\,\ol h'A(\D)v\cdot\id h\,,\cr
    S_L&=L\,\ol h^*F(\D)h - LJ\,\ol h'A(\D)P_+F(\D)h\,,\cr
    S_M&=K\,\ol h^*\Gamma\q - KJ\,\ol h'A(\D)P_+\Gamma\q+K'\,
         \ol\q\Gamma'h'\,.\cr}}
Note that the original source term $S_J$ cancels against the shift in
$\L_0$ in \shiftresult, but new terms appear in $S_L$ and $S_M$.
Replacing the dummy variable $\ol h^*$ by $\ol h$, we recover the
generating functional \genfun, but with the source $S_J$ changed,
\eqn\newsj{
    S_J\to-LJ\,\ol h'A(\D)P_+F(\D)h-KJ\,\ol h'A(\D)P_+\Gamma\q\,.}
We now repeat the derivative in eq.~\firstdef.  When we set the sources
to zero, we see that a derivative with respect to $L$ or $K$ must come
with a derivative with respect to $J$ to give a nonzero contribution.
We then find
\eqn\almostthere{\eqalign{
    \langle0|\,T\big\{\ol\q\Gamma'h'(z'),\,&\ol h'A(\D)v\cdot\id
    h(0),\, \ol hF(\D)h(x),\,\ol h\Gamma\q(z)\big\}|0\rangle\cr
    &=\im\delta(z)\,\langle0|\,T\big\{\ol\q\Gamma'h'(z'),\,
    \ol hA(\D)h(x),\,\ol h'A(\D)P_+\Gamma\q(0)\big\}|0\rangle\cr
    &\quad+\im\delta(x)\,\langle0|\,T\big\{\ol\q\Gamma'h'(z'),\,
    \ol h'A(\D)P_+F(\D)h(0),\,\ol h\Gamma\q(z)\big\}|0\rangle\,.\cr}}

Finally, we must perform the integral $(\im\int\d x\int\d z\d z'
\e^{\im k\cdot z}\e^{\im k'\cdot z'}v\cdot\id_z v'\cdot\id_{z'})$
to recover the matrix element \lszapply.  In this integral, the first
term on the right-hand side of eq.~\almostthere\ vanishes for an
on-shell state with $v\cdot k=0$, because the integral over $z$ is
trivial and there is no longer a one-particle pole to pick out.  The
second term, however, yields an $S$-matrix element in the usual way,
\eqn\smatrix{\eqalign{
    \im\int\d x\,&\im\delta(x)\int\d z\d z'\e^{\im k\cdot z}
    \e^{\im k'\cdot z'}v\cdot\id_z\,v'\cdot\id_{z'}\cr
    &\cdot\langle0|\,T\big\{\ol\q\Gamma'h'(z'),\,
    \ol h'A(\D)P_+F(\D)h(0),\,\ol h\Gamma\q(z)\big\}|0\rangle\cr
    &\qquad\qquad\qquad\qquad\qquad\qquad\qquad
    =-\langle M(v')|\,\ol h'A(\D) P_+ F(\D) h(0)\,|M(v)\rangle\,.}}
Thus we obtain the desired identity,
\eqn\again{\eqalign{
    \im\int\d x\,\langle M(v')|\,T\big\{\ol h'A(\D)v\cdot\id h,\,
    &\ol h F(\D)h(x)\big\}|M(v)\rangle\cr
    =&-\langle M(v')|\,\ol h'A(\D) P_+ F(\D) h\,|M(v)\rangle\,.\cr}}
Note that the term in $F(\D)$ proportional to $(v\cdot\id)^2$ will
not contribute here, since this matrix element is of the form
\firstbye\ and hence vanishes by the equation of motion.

A few additional comments are in order.  First, Politzer's result
\firstbye\ for the matrix elements of an operator which vanishes by the
equation of motion follows from an identical derivation, but with the
derivative $\delta/\delta\im L$ omitted. In this case the second term
of eq.~\almostthere\ does not appear, and we obtain zero instead of the
right-hand side of eq.~\smatrix.  (We stress that the intermediate
result \almostthere, which is the key to both proofs, is derived by
Politzer in full generality.)  Second, it is clear how this result is
to be generalised to the time-ordered product of an arbitrary number of
operators.  Essentially, we obtain a term on the right-hand side for
each contraction of $\ol h' A(\D)v\cdot\id h$ with an operator
insertion $\ol hG(\D)h$, where $G(\D)$ is any function of covariant
derivatives.  More than one contraction may be required to give a
nonzero result; for example, an operator of the form $\ol
h'A(\D)(v\cdot\id)^nh$ will have a nonvanishing matrix element only
when included in a time-ordered product with $n$ other operators such
as $\ol hF(\D)h$.  Finally, we reiterate that while for concreteness we
have framed our derivation within the heavy quark effective theory, it
is in fact completely general.

\listrefs
\listfigs

\bye